\pdfoutput=1
\PassOptionsToPackage{unicode}{hyperref}
\PassOptionsToPackage{hyphens}{url}
\PassOptionsToPackage{dvipsnames,svgnames,x11names}{xcolor}
\documentclass[
  british,
  11pt,
  a4paper,
]{article}
\usepackage{xcolor}
\usepackage[margin=25mm]{geometry}
\usepackage{amsmath,amssymb}
\usepackage{iftex}
\ifPDFTeX
  \usepackage[T1]{fontenc}
  \usepackage[utf8]{inputenc}
  \usepackage{textcomp} % provide euro and other symbols
\else % if luatex or xetex
  \usepackage{unicode-math} % this also loads fontspec
  \defaultfontfeatures{Scale=MatchLowercase}
  \defaultfontfeatures[\rmfamily]{Ligatures=TeX,Scale=1}
\fi
\usepackage{lmodern}
\ifPDFTeX\else
\fi
\IfFileExists{upquote.sty}{\usepackage{upquote}}{}
\IfFileExists{microtype.sty}{% use microtype if available
  \usepackage[]{microtype}
  \UseMicrotypeSet[protrusion]{basicmath} % disable protrusion for tt fonts
}{}
\makeatletter
\@ifundefined{KOMAClassName}{% if non-KOMA class
  \IfFileExists{parskip.sty}{%
    \usepackage{parskip}
  }{% else
    \setlength{\parindent}{0pt}
    \setlength{\parskip}{6pt plus 2pt minus 1pt}}
}{% if KOMA class
  \KOMAoptions{parskip=half}}
\makeatother
\usepackage{longtable,booktabs,array}
\usepackage{calc} % for calculating minipage widths
\usepackage{etoolbox}
\makeatletter
\patchcmd\longtable{\par}{\if@noskipsec\mbox{}\fi\par}{}{}
\makeatother
\IfFileExists{footnotehyper.sty}{\usepackage{footnotehyper}}{\usepackage{footnote}}
\makesavenoteenv{longtable}
\ifLuaTeX
\usepackage[bidi=basic,shorthands=off]{babel}
\else
\usepackage[bidi=default,shorthands=off]{babel}
\fi
\ifLuaTeX
  \usepackage{selnolig} % disable illegal ligatures
\fi
\providecommand{\tightlist}{%
  \setlength{\itemsep}{0pt}\setlength{\parskip}{0pt}}
\usepackage{microtype}
\usepackage{needspace}
\usepackage{setspace}
\usepackage{bookmark}
\IfFileExists{xurl.sty}{\usepackage{xurl}}{} % add URL line breaks if available
\hypersetup{
  pdftitle={Cited Web Sources in LLM Answers about Brands: A Multilingual Audit},
  pdfauthor={Dmitrij Żatuchin},
  pdflang={en-GB},
  colorlinks=true,
  linkcolor={Maroon},
  filecolor={Maroon},
  citecolor={Blue},
  urlcolor={Blue},
  pdfcreator={LaTeX via pandoc}}

\title{Cited Web Sources in LLM Answers about Brands:\\[4pt]A Multilingual Audit}
\author{Dmitrij Żatuchin}
\date{Revised preprint, 11 November 2026}

\begin{document}
\maketitle

\subsection{Abstract}\label{abstract}

Displayed citations make part of generative search observable, but they
do not reveal its complete retrieval process. This audit covers three
archived datasets containing 167,551 URL-bearing citation entries across
128 brands, 13 languages and 12 home markets. In the principal
Nordic-Baltic component, 3,863 of 21,033 registrable domains account for
80\% of 131,514 citations. Wikipedia leads in 11 of 12 query languages.
A recovered, separate brand-associated website classification retains
131,512 entries: 108,410 (82.4\%) fall outside its brand-associated
category and 23,102 (17.6\%) inside. This classification does not
establish legal ownership. In the full 57-brand Polish aggregate, four
recruitment hosts receive 637 citations against 297 for Polish
Wikipedia. The results show why normalization, classification rules and
denominators belong in citation audits. They describe displayed links
without estimating causal effects on brand reputation.

\textbf{Keywords:} citation analysis; generative search; information
sources; multilingual information access; domain concentration; large
language models

\subsection{1 Introduction}\label{introduction}

Generative search presents users with an answer and, in some
configurations, links to supporting web pages. The visible answer and
its citations are distinct objects of study. An answer can mention a
company without citing its website, cite several pages for a single
claim, or display a source whose relationship to the surrounding text is
uncertain. For information science, these links provide an observable
record of source presentation: which websites enter the answer
interface, how frequently they appear, and how that pattern varies
across query languages and systems.

Retrieval-augmented generation is one architecture for combining
retrieved information with language generation \textsuperscript{1}. Its
existence does not make an output citation a log of the complete
retrieval process. Systems can retrieve pages they do not cite, generate
material from model parameters, or attach citations after selecting an
answer. Establishing whether a passage supports a generated claim
requires a separate attribution assessment. Rashkin et al.~formalize
that distinction through attributable-to-identified-sources evaluation
\textsuperscript{2}. Liu et al.~show why it matters empirically:
displayed citations in generative search do not always provide adequate
support for the associated answer \textsuperscript{3}. The present study
counts citations without treating their presence as proof of support,
factual accuracy, or causal influence.

A further distinction concerns the brands named in an answer. The
synthesised consideration set describes the brands supplied by an AI
system; observed inclusion does not establish consumer consideration or
choice \textsuperscript{4}. The present audit examines the links
displayed with answers. A brand mention and a citation are therefore
separate measurement units.

Source presentation also matters independently of support. Li and
Sinnamon examine how generative search constructs authority and how
source patterns vary across systems and topics \textsuperscript{5}. Work
on news citations and on web search in generative systems extends this
agenda to outlet selection and differences between search interfaces
\textsuperscript{6,7}. These studies motivate audits of the distribution
of cited sources. Their samples, tasks and collection periods differ
from those examined here, so their findings provide context rather than
external validation of this corpus.

Brands provide a useful, bounded query domain. Questions about companies
can elicit reference works, corporate sites, news, marketplaces,
recruitment portals, videos and other sources. A multilingual sample
makes it possible to examine whether the leading cited domain is the
same in each query language. It also exposes practical measurement
problems: redirector URLs may conceal a publisher host, language
subdomains can split a single reference service into multiple units, and
a domain containing a brand name may not be legally owned by that brand.

This study brings together three existing research datasets, rather than
introducing a controlled experiment with a common sampling design. Their
combined inventory covers 128 distinct brand names, 13 languages and 12
home markets. The Nordic-Baltic component supports the main domain and
language analyses; the Polish component provides a separate aggregate
comparison; the Central and Eastern European component illustrates the
limits of combining URL citations with keyword-attributed source
mentions. The combined inventory describes coverage, not a common
denominator for every result.

The contribution is a reproducible descriptive audit of displayed source
distributions and an account of the measurement decisions needed to
interpret them. Four questions organize the analysis:

\begin{enumerate}
\def\labelenumi{\arabic{enumi}.}
\tightlist
\item
  How are citation entries and non-URL attributions distributed across
  the three components, and which comparisons do their units support?
\item
  How concentrated are cited domains in the Nordic-Baltic component, and
  how do leading domains vary by query language?
\item
  What source patterns appear in the separate Polish aggregate and
  across the recorded system labels?
\item
  How do redirectors, domain normalization and source-classification
  heuristics constrain interpretation?
\end{enumerate}

These are measurement questions. No hypothesis is tested about whether
publicity, website changes or citations improve reputation or
recommendations. Research on generative engine optimization studies
interventions under its own evaluation settings \textsuperscript{8}; the
observational counts reported here do not estimate the effects of such
interventions.

\subsection{2 Data and methods}\label{data-and-methods}

\subsubsection{2.1 Components and units}\label{components-and-units}

The three components were collected by the same research programme in
2026 using brand-oriented prompts and systems with web citation
functionality. They were assembled under related but non-identical
protocols. Table 1 distinguishes their native units. A citation entry
means a recorded occurrence, not a unique URL, document, answer or user
exposure. Repeated references remain in the count.

\Needspace{19\baselineskip}

Table 1. Corpus inventory. The combined URL-bearing total is an
inventory count; the main domain analysis uses the Nordic-Baltic
component.

{\def\LTcaptype{none} % do not increment counter
\begin{longtable}[]{@{}
  >{\raggedright\arraybackslash}p{(\linewidth - 6\tabcolsep) * \real{0.2500}}
  >{\raggedleft\arraybackslash}p{(\linewidth - 6\tabcolsep) * \real{0.2500}}
  >{\raggedleft\arraybackslash}p{(\linewidth - 6\tabcolsep) * \real{0.2500}}
  >{\raggedleft\arraybackslash}p{(\linewidth - 6\tabcolsep) * \real{0.2500}}@{}}
\toprule\noalign{}
\begin{minipage}[b]{\linewidth}\raggedright
Component
\end{minipage} & \begin{minipage}[b]{\linewidth}\raggedleft
Native records
\end{minipage} & \begin{minipage}[b]{\linewidth}\raggedleft
URL-bearing entries
\end{minipage} & \begin{minipage}[b]{\linewidth}\raggedleft
Brands
\end{minipage} \\
\midrule\noalign{}
\endhead
\bottomrule\noalign{}
\endlastfoot
Nordic-Baltic (NB) & 150,093 attribution rows & 131,514 & 66 \\
Poland (PL) & 4,151 answer records with citation arrays & 35,880 & 57 \\
Central and Eastern Europe (CEE) & 4,001 attribution rows & 157 & 20 \\
Combined & 189,974 attribution entries after expanding PL arrays &
167,551 & 128 distinct names \\
\end{longtable}
}

Brands overlap between components. The union of brand names is therefore
smaller than the sum of the component counts. NB covers 12 query
languages and brands with home markets in the Czech Republic, Denmark,
Estonia, Finland, Germany, Latvia, Lithuania, Norway, Poland, Slovakia
and Sweden. CEE adds Hungarian and Hungary to the combined coverage.
Czech codes \texttt{cs} and \texttt{cz} denote the same language when
counting overall coverage. Home market is a brand attribute, not a
statement about the user's physical location or the geographic origin of
every citation.

The NB file records brand, query language, language family, model label,
recognition tier, prompt identifier, home market, source type, citation
URL, title, snippet, attribution method and citation count. It does not
contain a complete collection-time or system-configuration manifest. The
PL file records brand, language, prompt, model, iteration and citation
arrays. CEE also contains keyword-derived source mentions. These exports
support finite-corpus summaries, but they cannot establish a balanced
factorial experiment or fully reconstruct the original querying
conditions.

\subsubsection{2.2 Analysis populations}\label{analysis-populations}

For NB, a URL-bearing row is one whose citation URL begins with
\texttt{http}. This gives 131,514 rows out of 150,093; the remaining
18,579 have no such URL. The domain analysis retains all 131,514
URL-bearing rows. A separate recovered classification excludes two rows
under its historical domain-parser rule and uses 131,512 entries, as
described in Section 2.4. The source-type contingency table uses all
150,093 attribution rows, including non-URL categories. It adds the
export label \texttt{implicit\_knowledge}, which is absent from
URL-bearing rows. This is a label on non-URL attributions, not an
observation of internal model knowledge. Other source categories also
contain non-URL attributions. Table 2 remains restricted to the ten
labels found on URL-bearing rows.

The PL inventory contains 35,880 citation entries across all 57 brands.
Its supplied resolved-domain aggregate also totals 35,880. Earlier
descriptions attached that total to a 46-brand national subset, but the
total belongs to the full component, which includes global benchmark
brands. The comparison reported here uses the complete supplied
aggregate. It does not represent an estimate for Polish national brands
alone.

CEE contains 157 URL-bearing entries and 3,844 other attributions,
predominantly keyword matches. A source name detected in answer prose is
not equivalent to a displayed URL. CEE therefore contributes to the
corpus inventory but not to NB concentration statistics, language
rankings or the PL portal comparison. The components are not pooled for
an overall source-type or ownership estimate.

\subsubsection{2.3 Redirector resolution and domain
normalization}\label{redirector-resolution-and-domain-normalization}

A total of 23,027 NB URL rows contain the host
\texttt{vertexaisearch.cloud.google.com}. For these rows, the export's
citation-title field contains the source host. The audit substitutes
this recorded host before counting domains. This is a transformation of
exported metadata, not a live visit proving that a redirect currently
resolves to the same page. No independent manual validation of these
substitutions is claimed.

For other NB URLs, the audit parses the hostname, lowercases it and
removes a leading \texttt{www.} only. It retains subdomains at the host
level. A literal \texttt{www.} sequence inside a hostname is preserved.
Ordinary \texttt{google.com} URLs are retained as Google URLs; they are
not automatically interpreted as redirectors. These choices avoid
corrupting legitimate hostnames or substituting missing titles for
ordinary Google pages.

The second aggregation uses registrable domains derived with
\texttt{tldextract} 5.3.1 and its bundled public-suffix snapshot, with
network retrieval disabled and private suffixes at the library default.
For example, Wikipedia language subdomains aggregate to
\texttt{wikipedia.org}; \texttt{rekvizitai.vz.lt} aggregates to
\texttt{vz.lt}. This normalization yields 23,399 host strings and 21,033
registrable-domain strings. These are computational grouping units, not
independently verified website operators or publishers. Registrable
domains can group heterogeneous services, whereas a publisher can
operate several domains.

Normalization changes exact domain counts. Earlier ledger calculations
used a last-label approximation, with a small set of multi-label suffix
exceptions. This revision uses a consistent public-suffix procedure and
reports the recomputed values throughout. Domain counts from
incompatible normalizations are not combined. The main rankings and
rounded rank-regression coefficients remain stable under the revised
procedure.

The PL comparison uses the supplied resolved aggregate rather than
claiming a new end-to-end reconstruction of its redirector mapping. Its
host labels and counts are preserved. Thus, the PL contrast between
several recruitment hosts and the Polish Wikipedia host is explicitly
different from the registrable-domain aggregation used for NB.

\subsubsection{2.4 Source labels and
ownership}\label{source-labels-and-ownership}

NB provides ten automated source-type labels, listed in Table 2. These
labels describe the classification output. They have not been
established as an exhaustive or independently validated taxonomy of
publisher types.

The source ledger reports a stricter brand-token or alias-based
ownership heuristic. Its exact alias dictionary is absent from the
public deposit, and its earlier 85.7\% outside / 14.3\% owned split has
not been reproduced. The present version withdraws that ownership
interpretation. A domain can contain a brand token without belonging to
the brand, while a subsidiary or country site can lack that token.

A different, archived correction of the brand-associated website
classifier was recovered and rerun for this version. It starts with the
source-type labels already stored in the NB export. For ordinary URLs,
it preserves the archived host parser. For the known Gemini Vertex
redirector, it extracts a registrable domain from the stored
citation-title field using an offline public-suffix snapshot. It then
identifies the modal pre-labelled company-site domain for each brand and
promotes a Gemini \texttt{other\_web} entry to \texttt{company\_website}
when its recovered domain equals that brand's inferred domain. Existing
labels for all other entries remain unchanged. This reproduces the
archived correction from the deposited labels; it does not independently
reconstruct or validate the initial labels.

The historical parser excludes two URL rows, at zero-based raw-file
indices 26,891 and 83,513, because their titles are \texttt{www.gov.pl},
which becomes a suffix-only string after normalization. They remain in
the full domain analysis and have excluded status in this
classification. Among the 131,512 retained entries, the correction
promotes 1,476 Gemini labels. The recovered row indices and source-type
labels match the saved historical output exactly. Five normalized domain
strings differ under the current bundled public-suffix list, with no
effect on the labels or counts.

The resulting \texttt{company\_website} category is reported here as
\textbf{brand-associated website classification}. Its inherited and
inferred domain matches can include company, group, country or
dealership sites; they are broader than verified legal ownership. A
single inferred primary domain can also miss other properties associated
with the brand. The output therefore supports a reproducible classifier
proportion, with unresolved classification error. It does not recover
the separate strict ownership rule. The ancillary script publishes the
recovered correction and its checks alongside the domain audit.

No manually annotated validation sample, inter-rater reliability result,
ownership registry check or citation-entailment assessment is available
for this study. A zero match or zero label must therefore not be
interpreted as evidence that a system never cites a brand's own site.
The analysis treats classification uncertainty as part of the
measurement problem. Repeated citation entries remain separately counted
in both populations.

\subsubsection{2.5 Summaries and statistical
interpretation}\label{summaries-and-statistical-interpretation}

Repeated-query audit protocols treat runs as repeated observations
within a question and system configuration and plan further sampling
from pilot estimates, without a universal repeat count
\textsuperscript{9}. The available exports do not retain enough
information to reconstruct all answer-level clusters. This audit
therefore reports descriptive summaries.

Domain concentration is calculated by sorting citation frequencies in
descending order and finding the smallest number of domains whose
cumulative count reaches the specified fraction of all NB URL entries.
Counts are occurrence-weighted: an answer with many citations
contributes more observations than an answer with few citations. A
language's leading domain is the domain with the highest count within
that query-language subset; its share uses that subset's URL count as
denominator.

For descriptive comparison with the earlier analysis, ordinary least
squares regresses log host frequency on log rank over the 1,000 most
frequently cited hosts. The negative slope and coefficient of
determination summarize that finite rank window. They do not constitute
a fitted probability distribution or a test distinguishing a power law
from alternatives. No tail threshold selection, distributional
goodness-of-fit test, likelihood comparison or population power-law
claim is made.

The source-type by language table is summarized with Pearson's
chi-square statistic and Cramér's V. Its entries are correlated: several
citations can belong to one answer, repeated prompts share brands and
systems, and source-level attribution can attach a citation to multiple
named entities. The row count therefore exceeds the number of
independent information-generation events. The chi-square statistic and
V are reported descriptively; the conventional row-independence p-value
is not used as evidence of a language effect. No cluster-adjusted
inference is claimed.

\subsubsection{2.6 Verification and reproducibility
scope}\label{verification-and-reproducibility-scope}

The revision audit rereads the deposited NB table, PL raw citation
inventory, PL resolved aggregate and CEE table. It checks row counts,
distinct brands, category counts, leading domains, language
denominators, concentration and the contingency statistic. Its script
records input hashes and the domain-normalization procedure. This
verifies computations against exported files, not the fidelity of the
original collection service or the content of cited pages.

The public deposit contains the analysis inputs and an earlier
reproduction script. This version adds ancillary code for the corrected
domain summaries, raw source-label summaries and recovered
brand-associated classification. The code checks the NB input hash,
states the separate denominators and records the two excluded
classification rows. It does not recover the missing strict ownership
dictionary or a complete collection-time configuration manifest. Data
and code availability are stated at the end of this paper.

\subsection{3 Results}\label{results}

\subsubsection{3.1 Recorded labels and recovered brand-associated
classification}\label{recorded-labels-and-recovered-brand-associated-classification}

The ten labels on NB URL rows sum to 131,514 (Table 2). The residual
\texttt{other\_web} category contains 101,285 entries (77.0\%);
\texttt{company\_website} contains 21,626 (16.4\%). These are
classification proportions on the same URL denominator. They are not
estimates of mutually validated publisher ownership categories. In
particular, the residual group may contain news, review and other
websites that the classifier did not assign to a narrower type.

\Needspace{19\baselineskip}

Table 2. Recorded source-type labels on NB URL rows. Rounding means
percentages need not sum exactly to 100\%.

{\def\LTcaptype{none} % do not increment counter
\begin{longtable}[]{@{}lrr@{}}
\toprule\noalign{}
Recorded label & Count & Share (\%) \\
\midrule\noalign{}
\endhead
\bottomrule\noalign{}
\endlastfoot
other\_web & 101,285 & 77.02 \\
company\_website & 21,626 & 16.44 \\
wikipedia & 5,192 & 3.95 \\
industry\_report & 1,301 & 0.99 \\
tier1\_news & 907 & 0.69 \\
social\_media & 563 & 0.43 \\
review\_platform & 394 & 0.30 \\
government & 192 & 0.15 \\
financial\_directory & 38 & 0.03 \\
academic & 16 & 0.01 \\
\end{longtable}
}

Domain counts and type labels are not interchangeable. The domain
procedure finds 5,800 Wikipedia citations, whereas the recorded
\texttt{wikipedia} label identifies 5,192. For Gemini, one of 23,032 URL
entries initially carries the \texttt{company\_website} label, a share
that rounds to 0.00\%, even though redirector substitution reveals
ordinary publisher and corporate domains. The labels were therefore not
consistently updated to reflect the same resolved source representation
used in the domain analysis. Rankings below rely on domains, not on
inferring domains from source-type labels.

The recovered correction produces 23,102 brand-associated website labels
among 131,512 retained citation entries (17.5665\%); the other
categories together contain 108,410 entries (82.4335\%). Table 3 reports
this result separately from the raw labels in Table 2. It covers the
same 66 brands, 12 languages and three system families. Repeated
citation entries remain counted, and the overall share is
citation-weighted.

\Needspace{15\baselineskip}

Table 3. Recovered brand-associated website classification. The
denominator excludes two rows under the archived parser and differs from
the full domain-analysis population.

{\def\LTcaptype{none} % do not increment counter
\begin{longtable}[]{@{}
  >{\raggedright\arraybackslash}p{(\linewidth - 6\tabcolsep) * \real{0.2500}}
  >{\raggedleft\arraybackslash}p{(\linewidth - 6\tabcolsep) * \real{0.2500}}
  >{\raggedleft\arraybackslash}p{(\linewidth - 6\tabcolsep) * \real{0.2500}}
  >{\raggedleft\arraybackslash}p{(\linewidth - 6\tabcolsep) * \real{0.2500}}@{}}
\toprule\noalign{}
\begin{minipage}[b]{\linewidth}\raggedright
Recorded system
\end{minipage} & \begin{minipage}[b]{\linewidth}\raggedleft
Retained entries
\end{minipage} & \begin{minipage}[b]{\linewidth}\raggedleft
Brand-associated
\end{minipage} & \begin{minipage}[b]{\linewidth}\raggedleft
Outside category
\end{minipage} \\
\midrule\noalign{}
\endhead
\bottomrule\noalign{}
\endlastfoot
gemini & 23,030 & 1,477 & 21,553 \\
openai & 18,206 & 2,954 & 15,252 \\
perplexity & 90,276 & 18,671 & 71,605 \\
Total & 131,512 & 23,102 (17.6\%) & 108,410 (82.4\%) \\
\end{longtable}
}

The 82.4\% complement includes Wikipedia and the other source categories
as well as the residual web category. It cannot be interpreted as an
earned-media share or a verified third-party ownership share. Assigning
both excluded rows inside versus outside the category gives 82.4323\% to
82.4338\% on the full URL denominator; both round to 82.4\%. This narrow
range addresses the two exclusions, not classification error.

Across all 150,093 NB attribution rows, the 12-language by
11-source-type table produces chi-square = 1,906.44 with 110 degrees of
freedom and Cramér's V = 0.0356. All expected cell counts exceed five.
The small V describes weak association in this table, but neither it nor
the large chi-square statistic corrects for within-answer dependence or
validates the labels. This result cannot establish that language has a
negligible causal effect on sourcing.

\subsubsection{3.2 Citation concentration}\label{citation-concentration}

On the registrable-domain basis, 3,863 of 21,033 domains (18.37\%)
account for at least 80\% of the 131,514 NB URL entries. The remaining
17,170 domains collectively account for approximately the other fifth.
The corresponding host-level threshold is 4,906 of 23,399 hosts
(20.97\%). Half of the citations are covered by 547 hosts. Reporting
both groupings makes the dependence of concentration on the analysis
unit visible.

The log-rank regression over the leading 1,000 hosts gives a slope of
-0.860 and R-squared of 0.983. It provides a compact description of a
steep, approximately linear relationship over this selected log-log
window. A high R-squared for ranked frequencies does not show that the
citation-generating process follows Zipf's law. The cumulative
thresholds provide the direct evidence for concentration without
requiring that stronger distributional interpretation.

\Needspace{19\baselineskip}

Table 4. Ten most frequently cited NB registrable domains, after
redirector substitution and normalization.

{\def\LTcaptype{none} % do not increment counter
\begin{longtable}[]{@{}lr@{}}
\toprule\noalign{}
Domain & Citation entries \\
\midrule\noalign{}
\endhead
\bottomrule\noalign{}
\endlastfoot
wikipedia.org & 5,800 \\
youtube.com & 2,826 \\
statista.com & 1,310 \\
wise.com & 908 \\
tatrabanka.sk & 733 \\
eset.com & 697 \\
reddit.com & 619 \\
europa.eu & 600 \\
printful.com & 599 \\
pipedrive.com & 592 \\
\end{longtable}
}

The head includes reference, video and statistics services alongside
corporate and institutional domains. These labels are interpretive
descriptions of familiar examples, not a newly validated source
taxonomy. The result is about recorded frequency, not source
credibility, readership or contribution to an answer's content.

\subsubsection{3.3 Leading domains by query
language}\label{leading-domains-by-query-language}

Wikipedia is the leading registrable domain in 11 of the 12 NB language
subsets. Its leading shares range from 3.71\% in Polish to 5.70\% in
Finnish. Lithuanian is the exception: \texttt{vz.lt} accounts for 463 of
10,576 URL entries (4.38\%). The domain includes the
\texttt{rekvizitai.vz.lt} subdomain, so this result concerns the
combined domain group rather than citations to the business newspaper
alone.

\Needspace{18\baselineskip}

Table 5. Leading registrable domain within each NB query language. All
denominators are URL-bearing citation occurrences.

{\def\LTcaptype{none} % do not increment counter
\begin{longtable}[]{@{}lrlrr@{}}
\toprule\noalign{}
Query language & URL entries & Leading domain & Count & Share (\%) \\
\midrule\noalign{}
\endhead
\bottomrule\noalign{}
\endlastfoot
Czech & 11,307 & wikipedia.org & 503 & 4.45 \\
Danish & 10,520 & wikipedia.org & 417 & 3.96 \\
German & 11,239 & wikipedia.org & 496 & 4.41 \\
English & 11,628 & wikipedia.org & 552 & 4.75 \\
Estonian & 10,608 & wikipedia.org & 464 & 4.37 \\
Finnish & 10,581 & wikipedia.org & 603 & 5.70 \\
Lithuanian & 10,576 & vz.lt & 463 & 4.38 \\
Latvian & 10,737 & wikipedia.org & 494 & 4.60 \\
Norwegian & 10,523 & wikipedia.org & 517 & 4.91 \\
Polish & 11,439 & wikipedia.org & 424 & 3.71 \\
Slovak & 11,310 & wikipedia.org & 427 & 3.78 \\
Swedish & 11,046 & wikipedia.org & 486 & 4.40 \\
\end{longtable}
}

These are descriptive differences across recorded query-language
subsets. The analysis does not measure the size or quality of each
language's web corpus and cannot explain the pattern as a fallback
caused by scarcity. Nor does a query-language difference identify a
market effect: brand mix, prompt realization, system configuration and
collection time may also contribute.

\subsubsection{3.4 The Polish aggregate}\label{the-polish-aggregate}

The supplied resolved PL aggregate contains 35,880 citations across 57
brands. YouTube is its most frequently cited host, with 2,289 entries
(6.38\% of the aggregate). Four recruitment-related hosts together
account for 637 entries (1.78\%), compared with 297 (0.83\%) for
\texttt{pl.wikipedia.org} (Table 6). Their count ratio is 2.14.

\Needspace{17\baselineskip}

Table 6. Selected hosts in the full 57-brand Polish aggregate. The
four-host total is compared with one language-specific Wikipedia host,
not all Wikipedia subdomains.

{\def\LTcaptype{none} % do not increment counter
\begin{longtable}[]{@{}lr@{}}
\toprule\noalign{}
Host or group & Citation entries \\
\midrule\noalign{}
\endhead
\bottomrule\noalign{}
\endlastfoot
pl.indeed.com & 176 \\
livecareer.pl & 169 \\
interviewme.pl & 159 \\
randstad.pl & 133 \\
Four recruitment hosts combined & 637 \\
pl.wikipedia.org & 297 \\
\end{longtable}
}

This is a selected group comparison. Four hosts have more opportunities
to accumulate citations than one host, and the selection is informed by
the observed corpus. The comparison does not establish dominance of
recruitment sites over reference sources generally. It identifies a
source group that merits attention in this particular audit. It supplies
no measurement of how users perceive the brands or how the cited
employment material affects recommendations.

\subsubsection{3.5 Differences across system
labels}\label{differences-across-system-labels}

The NB export identifies three system families as \texttt{perplexity},
\texttt{gemini} and \texttt{openai}. The earlier manuscript names Sonar
Pro, Gemini 3.1 Pro and GPT-5.4; those version claims are not
independently established by the exported row fields. Table 7 therefore
uses the recorded family labels. These refer to the sampled
configurations, not to all products or versions offered by the
providers.

\Needspace{14\baselineskip}

Table 7. Full NB URL entries and distinct registrable domains by
recorded system label. Wikipedia share is domain-derived. The separate
retained classification counts appear in Table 3.

{\def\LTcaptype{none} % do not increment counter
\begin{longtable}[]{@{}lrrr@{}}
\toprule\noalign{}
Recorded system & URL entries & Domains & Wikipedia (\%) \\
\midrule\noalign{}
\endhead
\bottomrule\noalign{}
\endlastfoot
perplexity & 90,276 & 16,158 & 4.89 \\
gemini & 23,032 & 6,609 & 2.64 \\
openai & 18,206 & 3,315 & 4.29 \\
\end{longtable}
}

Perplexity accounts for 68.64\% of NB URL entries and has the largest
observed set of domains. Larger citation volume increases the
opportunity to observe distinct domains, so the domain counts do not
establish greater diversity under equal sampling effort. The aggregate
is also heavily weighted toward this system's citations. An equal-system
or equal-answer summary could differ substantially.

These differences reflect the combined collection and presentation
systems, including any retrieval provider, API settings, prompt handling
and citation-display rules. Their effects cannot be separated from model
effects with the available exports. The change from one raw Gemini
company-site label to 1,477 after the recovered correction illustrates
the effect of preprocessing on a system comparison. It does not measure
a change in model knowledge.

\subsection{4 Discussion}\label{discussion}

\subsubsection{4.1 Implications for information-source
measurement}\label{implications-for-information-source-measurement}

The audit identifies a concentrated but extensive set of cited domains,
with a shared leading reference source in most query languages and a
different leader in Lithuanian. Its contribution lies in the observable
citation layer. A domain-frequency table can describe which source links
are presented to users even when the system's complete retrieval trace
is unavailable. That table becomes interpretable only when its row unit,
normalization and sampling weights are explicit.

The distinction between a host and a registrable domain is substantive.
Combining language subdomains makes Wikipedia comparable across
languages; combining the Lithuanian subdomains changes which domain
group leads. Neither grouping is universally correct. A study of
publishers, services or individual pages may require a different unit.
Reporting the choice allows readers to identify what a ranking can and
cannot answer.

The results also show why citation inventories and source-type
classifications should be kept separate. A link can be counted and
grouped even when its publisher category is uncertain. Conversely, a
keyword match can carry a source-type label without supplying any
displayed link. Pooling these forms of evidence would obscure the
distinction between observed links and inferred source mentions. The
very small URL-bearing share of CEE makes that risk visible.

\subsubsection{4.2 Implications for auditing
practice}\label{implications-for-auditing-practice}

An organization can use the leading-domain lists to inspect recurring
citations and check whether the linked material is accurate, current and
relevant to the answer. A manageable head of domains provides a
practical starting point, while the large tail cautions against assuming
that a short monitoring list is exhaustive. Auditing should retain the
original URL and answer context so that aggregate patterns can be traced
to individual instances.

Language-specific checks remain useful even when most leading domains
are shared. The Lithuanian result would be lost if the analysis
considered only English outputs, and the Polish aggregate points to
recruitment-related sources that a reference-only audit would omit.
These findings justify broader observation within the audited task
domain; they do not prescribe paid placement, editorial intervention or
a guaranteed route to improved answer visibility.

A further practical requirement is to separate technical failures from
substantive findings. Counting redirectors as publishers would change
the apparent leading sources. Using unresolved labels could produce a
false zero for company sites. A source audit should therefore inspect
transformations before interpreting a pattern as a property of brands or
systems.

\subsubsection{4.3 Limits of causal and comparative
interpretation}\label{limits-of-causal-and-comparative-interpretation}

Citation frequency cannot reveal the causal influence of a page on
generated text. That would require additional evidence, such as
controlled source interventions or system-level retrieval traces. It
also cannot establish reputational consequences without measuring
relevant users, judgments or outcomes. No such evidence was collected
here.

Likewise, the three components do not constitute independent
replications under a shared design. They differ in brand coverage,
available metadata and attribution procedure. Home-market coverage
describes the assembled corpus; it does not imply a controlled
comparison of 12 national populations. Within NB, language and system
summaries remain conditional on the sampled brands, prompts, periods and
configurations.

\subsection{5 Limitations and further
work}\label{limitations-and-further-work}

The first limitation is coverage. The brands and prompts form a
purposive research corpus, not a probability sample of brands, users or
web searches. The combined scale should not be read as
representativeness. A high number of citation entries can arise from
repeated answers or citation-rich responses without adding equivalent
independent information.

Second, the exports do not preserve a complete timestamped configuration
manifest. Exact versions, search backends, decoding parameters and
temporal changes therefore cannot be held constant analytically.
Recording these variables and repeated collection windows would permit
more informative future comparisons. The present results should remain
tied to the archived corpus.

Third, attribution occurs at source or response level rather than being
independently anchored to every brand-specific proposition. When an
answer includes multiple brands, the same citation may be associated
with more than one entity. Deduplicating answer-level events or
evaluating claim-specific support would answer different questions. A
cluster-aware design would also be needed before treating row-level
associations as inferential evidence.

Fourth, neither source-type labels nor ownership heuristics have been
validated against a study-specific human reference set. The omitted
strict ownership alias dictionary prevents reproduction of the earlier
ownership estimate. The recovered correction reproduces a separate
brand-associated classification from the deposited labels, with no
ownership adjudication. Future work should publish dated domain
mappings, define ambiguous group and dealership cases, and assess a
stratified sample without borrowing validation from another study.

Fifth, domain normalization groups some distinct services and separates
some common publishers. The public-suffix snapshot and exact parsing
rule are part of the method, and changing them may change counts. The
rank-regression summary is similarly conditional on its chosen window;
it provides no formal evidence for a power-law generating process.

Finally, the PL resolved aggregate supports its published host counts,
but the mapping from raw redirectors to every aggregate host has not
been independently reconstructed in this revision. Its claims are
therefore restricted to the supplied aggregate. The CEE keyword material
cannot remedy this limitation or validate the larger URL analyses.

\subsection{6 Conclusion}\label{conclusion}

The three archived components contain 167,551 URL-bearing citation
entries within 189,974 attribution rows. Their largest URL component,
NB, shows substantial domain concentration: 3,863 of 21,033 normalized
domains account for 80\% of 131,514 citations. Wikipedia leads in 11 of
12 query languages, with \texttt{vz.lt} leading in Lithuanian. The
recovered classification places 108,410 of 131,512 retained entries
(82.4\%) outside its brand-associated website category. In the separate
full Polish aggregate, four selected recruitment hosts receive 637
citations compared with 297 for Polish Wikipedia.

These results describe displayed citations in a defined corpus. They
also show that preprocessing, domain grouping, classifier limitations
and unequal citation volume materially affect interpretation. Citation
audits can support scrutiny of the sources visible in generative answers
when those choices are transparent. They cannot, on their own, establish
complete retrieval provenance, training sources, legal ownership or
causal effects on brand reputation.

\subsection{Declarations}\label{declarations}

\textbf{Author and affiliations.} Dmitrij Żatuchin, Department of
Information Technologies, Estonian Entrepreneurship University of
Applied Sciences (EUAS), Tallinn, Estonia; Rankfor.AI Sp. z o.o.,
Wrocław, Poland. ORCID: \url{https://orcid.org/0009-0002-1929-9770}.
Correspondence:
\href{mailto:dmitrij@rankfor.ai}{\nolinkurl{dmitrij@rankfor.ai}}.

\textbf{Funding and competing interests.} This research received no
external funding. The author is affiliated with Rankfor.AI Sp. z o.o.,
which develops AI brand-intelligence tools. The three datasets are
Rankfor.AI research data. The analysis and conclusions are the author's
own.

\textbf{Author contributions and assistance.} The author conceived the
study, designed the methodology, performed the citation classification
and analysis, and wrote the manuscript. OpenAI Codex assisted with
revision, computational checks and formatting. Automated assistance does
not supply human validation of citation or ownership classifications.

\textbf{Ethics.} The study analysed AI-system outputs and involved no
human participants. No ethics-committee approval or participant consent
is claimed.

\textbf{Data and code availability.} The existing data deposit is
available under CC BY 4.0 at
\url{https://doi.org/10.5281/zenodo.20829524}. It contains the NB and
CEE attribution tables, Polish citation arrays, resolved aggregates and
an earlier reproduction script. The present version supplies revised
audit code, instructions and dependency versions as ancillary files.
These reproduce summaries from the archived exports and labels. They do
not reconstruct the original collection, establish legal ownership or
validate whether a cited page supports an answer. The strict alias
dictionary used for the earlier 14.3\% ownership claim remains
unavailable.

\textbf{Version note.} This version revises arXiv:2606.25787v1. It
corrects domain counts and the scope of the Polish aggregate, replaces
the unreproduced 85.7\% ownership assertion with a separately reproduced
82.4\% brand-associated classification result, and narrows
interpretation to displayed citations. It also adds the published
marketing-analytics study and the corrected Dice Roll Method protocol as
related work. The original public data deposit is unchanged.

\subsection{References}\label{references}

\begin{enumerate}
\def\labelenumi{\arabic{enumi}.}
\tightlist
\item
  Lewis P, Perez E, Piktus A, et al.~Retrieval-augmented generation for
  knowledge-intensive NLP tasks. In: \emph{Advances in Neural
  Information Processing Systems}. 2020;33:9459-9474.
\item
  Rashkin H, Nikolaev V, Lamm M, et al.~Measuring attribution in natural
  language generation models. \emph{Computational Linguistics}.
  2023;49(4):777-840.
\item
  Liu NF, Zhang T and Liang P. Evaluating verifiability in generative
  search engines. In: \emph{Findings of the Association for
  Computational Linguistics: EMNLP 2023}. 2023, pp.~7001-7025.
  \url{https://doi.org/10.18653/v1/2023.findings-emnlp.467}
\item
  Żatuchin D. The synthesised consideration set: brand inclusion and
  concentration in large language model recommendations across five
  industries and three systems. \emph{Journal of Marketing Analytics}.
  Published online 31 October 2026.
  \url{https://doi.org/10.1057/s41270-026-00562-3}
\item
  Li A and Sinnamon L. Generative AI search engines as arbiters of
  public knowledge: An audit of bias and authority. \emph{Proceedings of
  the Association for Information Science and Technology}.
  2024;61(1):205-217. \url{https://doi.org/10.1002/pra2.1021}
\item
  Yang KC. News source citing patterns in AI search systems. Preprint,
  arXiv:2507.05301, 2025.
  \url{https://doi.org/10.48550/arXiv.2507.05301}
\item
  Kirsten E, Grosse Perdekamp J, Wu Q, et al.~Characterizing web search
  in the age of generative AI. Preprint, arXiv:2510.11560v2, 2025,
  revised 2026. \url{https://doi.org/10.48550/arXiv.2510.11560}
\item
  Aggarwal P, Murahari V, Rajpurohit T, et al.~GEO: Generative engine
  optimization. In: \emph{Proceedings of the 30th ACM SIGKDD Conference
  on Knowledge Discovery and Data Mining}. 2024, pp.~5-16.
  \url{https://doi.org/10.1145/3637528.3671900}
\item
  Żatuchin D. The Dice Roll Method: A Standardized Protocol for
  Repeated-Query Auditing of Large Language Model Brand Recommendations.
  Preprint, arXiv:2609.04047v2, revised 14 October 2026.
  \url{https://doi.org/10.48550/arXiv.2609.04047}
\end{enumerate}

\end{document}